\documentclass{article}
\usepackage[utf8]{inputenc}
\usepackage{graphicx,hyperref}

\title{Trade-offs between sustainable development goals in systems of cities}

\author{J. Raimbault$^{1,\ast}$ and D. Pumain$^{2}$\medskip\\
$^{1}$ CASA, University College London\\
$^{2}$ G{\'e}ographie-cit{\'e}s, Universit{\'e} Paris 1\medskip\\
$\ast$ \texttt{juste.raimbault@polytechnnique.edu}
}
\date{}

\begin{document}

\maketitle

\begin{abstract}
	Sustainable Development Goals are intrinsically competing, but their embedding into urban systems furthermore emphasises such compromises, due to spatial complexity, the non-optimal nature of such systems, and multi-objective aspects of their agents, among other reasons. We propose in this paper to use a stylised simulation model for systems of cities, focused on innovation diffusion and population dynamics, to show how trade-offs may operate at such a scale. We proceed in particular to a bi-objective optimisation of emissions and innovation utilities, and show that no single urban optimum exists, but a diversity of regimes forming a compromise between the two objectives.
\end{abstract}

\section{Introduction}

The concept of urban optima, in the sense of optimising certain dimensions of urban systems, has been considered from diverse perspectives. In most cases, there does not seem to be clear patterns, neither empirical nor theoretical, of possible simple optimisations of single objectives by urban systems. Some results in urban economics regarding an optimal city size requires to consider a city in a closed system, which is unreasonable from a realistic perspectives \cite{singell1974optimum}. Studies of an optimal urban population density are restricted to economic criteria of wage and productivity \cite{su2017density}. The sustainability of urban forms for CO2 emissions requires considering complex indicators of urban form \cite{le2012urban}. In terms of pollution, empirical results across different urban systems suggest no fixed relationship between city size and emission of pollutants \cite{han2016optimum}. Altogether, this converges with the idea of multiple agents optimising multiple dimensions at different scales \cite{pumain2008socio}.

Sustainable Development Goals (SDGs) are characterised in a similar way by compromises between different dimensions. Urban sustainability, in the sense of the urban aspect of environmental issues \cite{finco2001pathways}, has thus to be understood as trade-offs between multiple objectives \cite{viguie2012trade}. This aspect occurs within subsystems themselves, such as in the case of designing transport networks \cite{sharma2011multiobjective}. Planning and policies must in that context account for such competing objectives \cite{caparros2015optimised}.

We propose in this paper to study trade-offs between different SDGs in systems of cities. We consider systems of cities at the macroscopic scale, and more particularly the dynamics of innovation diffusion and population growth. Using a stylised model for such urban dynamics, we apply a bi-objective optimisation genetic algorithm, to explore how trade-offs can occur in such systems. The rest of this paper is organised as follows: we first recall the assumptions of the system of cities model applied; we then describe results of its optimisation on proxies for two SDGs (innovation utility and emissions); we finally discuss theoretical implications of these results and how further work could include empirical components.

\section{Urban system model}

We work with a stylised model for the dynamics of urban systems at the macroscopic scale. This model is based on innovation diffusion dynamics and their impact on population growth. It was first formulated by \cite{favaro2011gibrat}, within the context of an evolutionary urban theory \cite{pumain1997pour}. A similar agent-based model was used to explore assumptions on the emergence of systems of cities themselves \cite{schmitt2015half}. A modified version was described by \cite{raimbault2020model} as an urban evolution model, including an urban genome shared and mutated across cities. As this particular version can furthermore be setup on stylised systems of cities, we use it in our multi-objective optimisation approach.

We describe the model below, but without detailed equations for the purpose of staying concise. See \cite{raimbault2020model} for the full description.

\subsection{Model setup}

The simulated urban system is composed by cities, characterised at each time step by (i) their population; (ii) their location in the geographical space; (iii) adoption rates by their populations for different innovations (urban genome).

We work on synthetic systems of cities, which are randomly generated given some fixed macro characteristics. This approach allows controlling for example for the role of space, and unentangling intrinsic model dynamics from geographical contingencies \cite{raimbault2019space}. In our case, as emission indicator is linked to inter-city flows, strongly dependent on the geography, averaging over several synthetic systems of cities will thus provide robust results.

Synthetic systems of cities are generated with random locations, an initial rank-size hierarchy which can be tuned (otherwise fixed to a default value of 1, to mirror a Zipf law distribution for city size), and a number of 30 cities. The largest city has initially a population of 100,000 and the model is run for 50 time steps.

\subsection{Model dynamics}

Starting from the initial state, the model updates population and innovation step by step. At each time step (of an order of magnitude of 10 years - the effects are observed on long time scales), the following procedure is used:

\begin{enumerate}
	\item innovations are diffused between cities using a spatial interaction model; innovations with a higher utility will diffuse more;
	\item populations are updated following an other spatial interaction model, with an advantage for cities being more innovative;
	\item new innovations may be invented in cities, following a probability determined by a mutation rate and by population (with a given hierarchy across cities);
	\item if a new innovation emerges, it has an initial penetration share fixed by one parameter, and a utility randomly distributed (normal or log-normal law), with a fixed standard deviation and an average corresponding to the current empirical average of existing innovation utilities.
\end{enumerate}

\subsection{Model parameters}

The parameters left free for optimisation are (i) spatial interaction range for innovation diffusion; (ii) spatial interaction range for population growth; (iii) mutation (innovation) rate; (iv) innovative city selection hierarchy; (v) rate of early adopters; (vi) standard-deviation of new innovation utilities; (vii) type of distribution for new innovation utilities.

\subsection{Optimisation objectives}

We consider the ``innovation'' SDG (goal 9) and the ``climate'' SDG (goal 14) as conflicting objectives. We can expect that a higher economic activity linked to more intensive innovative activities will increase endogenous emissions, but also transport emissions between urban areas, generated by economic and transport flows. The existence of a trade-off remains an hypothesis which will be checked during the optimisation stage.

We consider therefore the two following objectives for model optimisation: (i) aggregated total utility during model dynamics, computed over time and across cities, with shares of each innovations; (ii) total emissions due to transport flows between cities, computed as cumulative population gravity flows.

\section{Results}

\subsection{Implementation}

The model is implemented in \texttt{scala} for performances purposes, using matrix operations to update innovation shares and populations. Source code and simulation results are available on the open git repository of the project at \url{https://github.com/JusteRaimbault/UrbanEvolution}.

Model optimisation is achieved by integrating the model into the OpenMOLE platform \cite{reuillon2013openmole}. This free and open source software facilitates model embedding into a workflow system, distribution of computation into high-performance computing infrastructures, and provides a simple access to state-of-the-art model validation methods.

\subsection{Bi-objective optimisation}

We investigate trade-offs between total innovation utility and emissions, by optimising the model using a bi-objective heuristic with free parameters and indicators detailed above. We use a NSGA2 optimisation algorithm, provided by OpenMOLE, with a population of 100 individuals, for 10,000 generations.

\begin{figure}
	\centering
	\includegraphics[width=\linewidth]{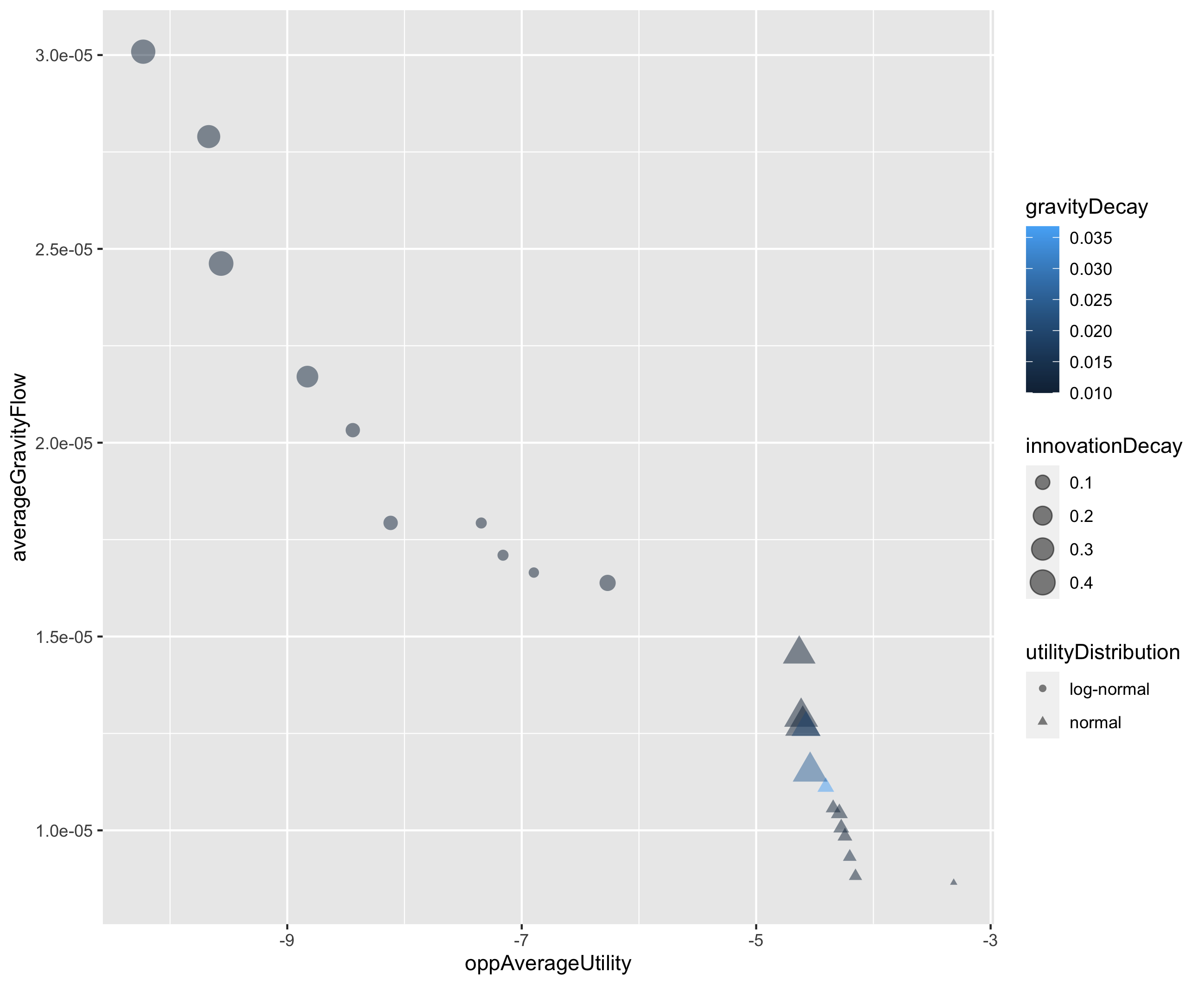}
	\caption{Pareto front between the opposite of average utility and cumulated gravity flows. Point color gives population spatial interaction range; point size innovation diffusion range; and point shape the utility distribution.\label{fig:fig1}}
\end{figure}

We show optimisation results, as the final algorithm population, in Fig.~\ref{fig:fig1}. We indeed find a broad Pareto front, confirming the existence of a trade-off in such urban dynamics driven by innovation diffusion. We note two parts of the Pareto front, with fat-tailed distributions for utility distribution (log-normal) giving the upper part of the front corresponding to situations with a higher utility but which are more emission intensive. Within this subfront, population spatial interaction are rather local, while a more local innovation diffusion yields less emitting configurations. A similar aspect is observed for the normal distribution subfront, with a U-shaped value of population spatial interactions when going through the front: a more integrated system in terms of population migration produces by itself an intermediate compromise.

\subsection{Conditional optimisation}


\begin{figure}
	\centering
	\includegraphics[width=\linewidth]{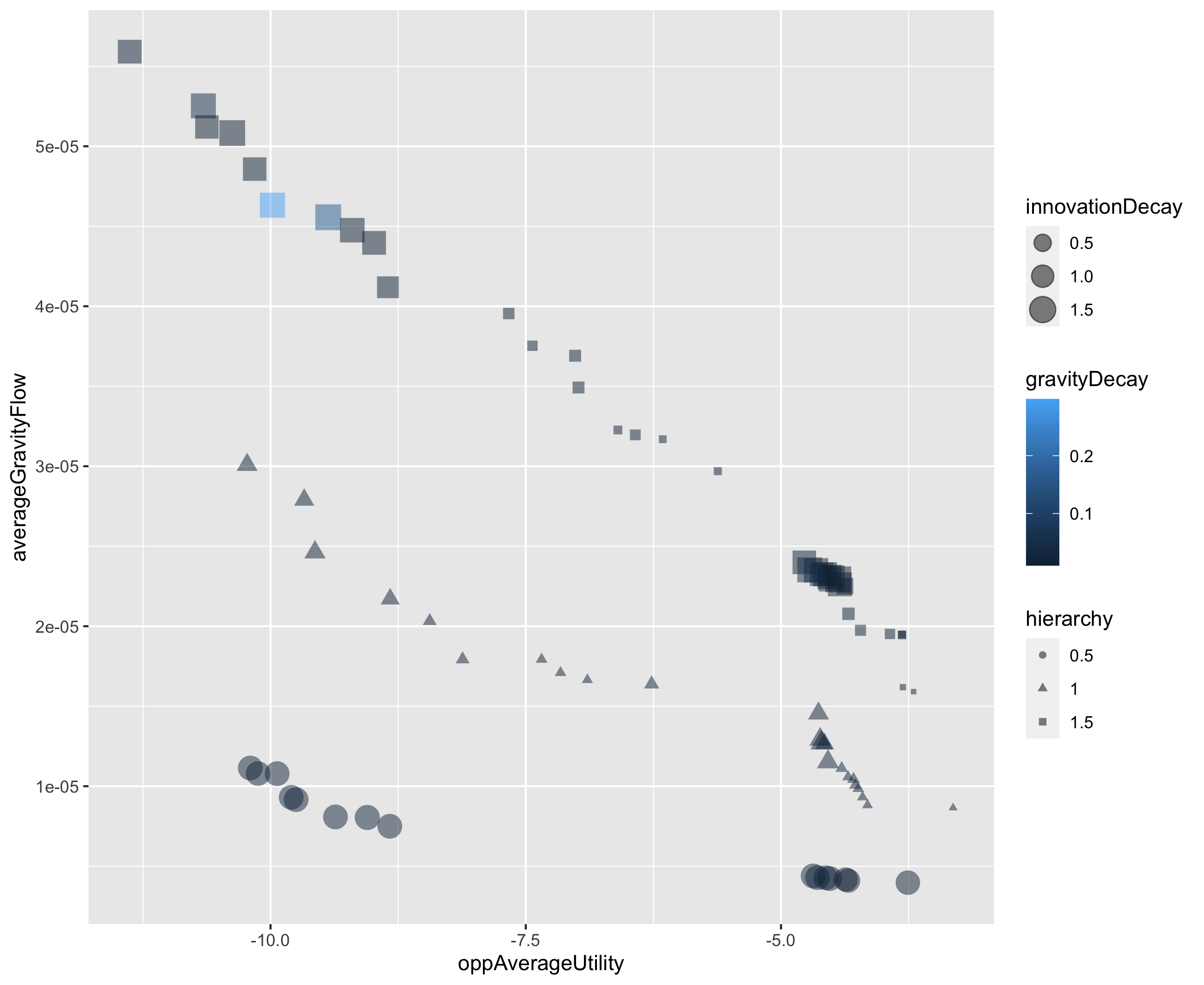}
	\caption{Pareto fronts, with initial population hierarchy fixed at different values (point shape).\label{fig:fig2}}
\end{figure}

We now turn to experiments which could potentially provide policy insights. We run the same optimisation as before, but changing the initial population hierarchy of cities. To put it simply, we investigate how trade-offs change in different hypothetical systems of cities, ranging from highly hierarchical (Zipf exponent of 1.5) to a more balanced system (exponent of 0.5). We show results in Fig.~\ref{fig:fig2}. We find that the higher the hierarchy, the less flat the front. Overall, the less hierarchical system dominates the others (but this comparison remains limited as total population is different across systems). Furthermore, the size of the front is the smallest with the lowest hierarchy, meaning that this system is indeed closer to some global optimum.

\begin{figure}
	\centering
	\includegraphics[width=\linewidth]{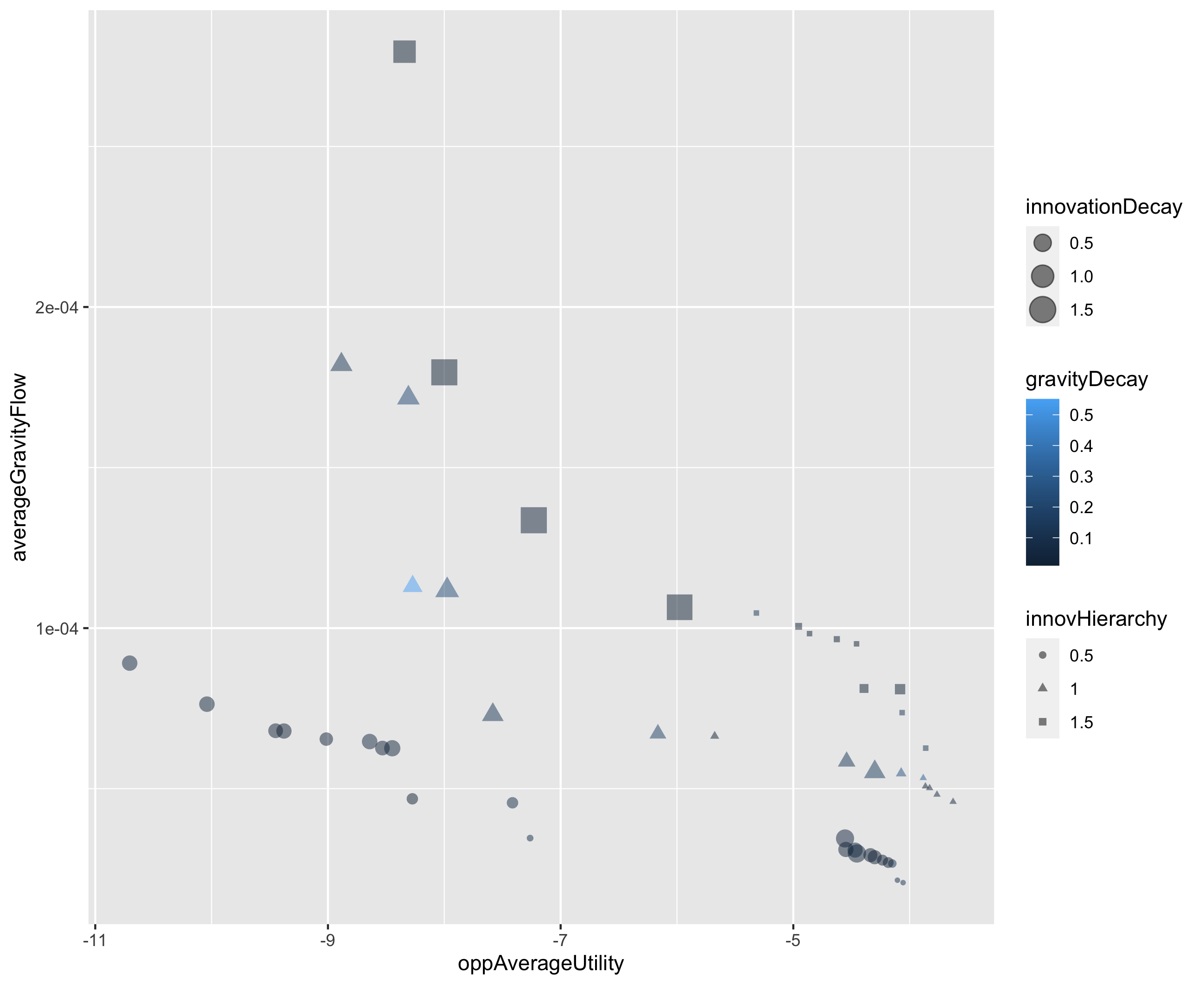}
	\caption{Pareto fronts, with innovation hierarchy fixed at different values (point shape).\label{fig:fig3}}
\end{figure}

We finally show in Fig.~\ref{fig:fig3} a similar conditional optimisation, run by changing the fixed value of innovation hierarchy. This corresponds in terms of policies, to either letting innovation aggregate into larger metropolises (scaling with a high exponent value \cite{pumain2006evolutionary}), or regulating and providing incentives to enhance innovation into smaller and medium-sized cities. We also find that balanced policies provides a more optimal front (they can be compared in this case). Furthermore, this lowest hierarchy corresponds to much higher absolute values of total utility, going against the narrative of a higher value innovation produced by large cities only. Points for the two other fronts are rather close, corresponding to a lower sensitivity when the scaling exponent is larger than one.

\section{Discussion}

We have shown, in a stylised model of urban population and innovation dynamics, that trade-offs between transport emissions and total innovation utility emerge from model dynamics. This has theoretical implications, confirming the general non-optimising nature of urban systems and the predominance of trade-offs across different urban dimensions. Our results from conditional optimisation suggest that less hierarchical systems, both regarding initial population hierarchy and innovation hierarchy, provide more optimal Pareto fronts. This could have implications for policies such as innovation incentives, to avoid a too strong concentration into larger cities. More empirical investigations remain however needed.

Extending this theoretical and stylised work towards more empirical and data-grounded applications raises several issues. First, how to quantify spatial proxies for innovation, to either parametrise the initial configuration, or to calibrate model trajectories in terms of innovation diffusion, remains a difficult question. The use of patent data provides such insights \cite{griliches200713}, but the lack of harmonised and spatialised open patent database limits this perspective. Some initiatives are currently working towards this goal, such as \cite{bergeaud2021patentcity}. Furthermore, more realistic indicators for emissions, both transport and endogenous ones, would be also needed, for example by estimating them through a link with existing emissions databases such as EDGAR \cite{olivier1994emission}.

To conclude, we have provided a first stylised insight into trade-offs between SDGs in systems of cities at the macroscopic scale, which can be applied from a theoretical viewpoint to validate or unvalidate urban theories, and be used as a basis towards more practical application towards sustainable long-term territorial policies \cite{rozenblat2018conclusion}.


\newpage

\begin{thebibliography}{}

\bibitem[Bergeaud and Verluise, 2021]{bergeaud2021patentcity}
Bergeaud, A. and Verluise, C. (2021).
\newblock Patentcity: A century of innovation: New data and facts.

\bibitem[Caparros-Midwood et~al., 2015]{caparros2015optimised}
Caparros-Midwood, D., Barr, S., and Dawson, R. (2015).
\newblock Optimised spatial planning to meet long term urban sustainability
  objectives.
\newblock {\em Computers, Environment and Urban Systems}, 54:154--164.

\bibitem[Favaro and Pumain, 2011]{favaro2011gibrat}
Favaro, J.-M. and Pumain, D. (2011).
\newblock Gibrat revisited: An urban growth model incorporating spatial
  interaction and innovation cycles.
\newblock {\em Geographical Analysis}, 43(3):261--286.

\bibitem[Finco and Nijkamp, 2001]{finco2001pathways}
Finco, A. and Nijkamp, P. (2001).
\newblock Pathways to urban sustainability.
\newblock {\em Journal of Environmental Policy \& Planning}, 3(4):289--302.

\bibitem[Griliches, 2007]{griliches200713}
Griliches, Z. (2007).
\newblock {\em 13. Patent Statistics as Economic Indicators: A Survey}.
\newblock University of Chicago Press.

\bibitem[Han et~al., 2016]{han2016optimum}
Han, L., Zhou, W., Pickett, S.~T., Li, W., and Li, L. (2016).
\newblock An optimum city size? the scaling relationship for urban population
  and fine particulate (pm2. 5) concentration.
\newblock {\em Environmental Pollution}, 208:96--101.

\bibitem[Le~N{\'e}chet, 2012]{le2012urban}
Le~N{\'e}chet, F. (2012).
\newblock Urban spatial structure, daily mobility and energy consumption: a
  study of 34 european cities.
\newblock {\em Cybergeo: European Journal of Geography}.

\bibitem[Olivier et~al., 1994]{olivier1994emission}
Olivier, J., Bouwman, A., Van~der Maas, C., and Berdowski, J. (1994).
\newblock Emission database for global atmospheric research (edgar).
\newblock {\em Environmental Monitoring and Assessment}, 31(1):93--106.

\bibitem[Pumain, 1997]{pumain1997pour}
Pumain, D. (1997).
\newblock Pour une th{\'e}orie {\'e}volutive des villes.
\newblock {\em L'Espace g{\'e}ographique}, pages 119--134.

\bibitem[Pumain, 2008]{pumain2008socio}
Pumain, D. (2008).
\newblock The socio-spatial dynamics of systems of cities and innovation
  processes: a multi-level model.
\newblock In {\em The Dynamics of Complex Urban Systems}, pages 373--389.
  Springer.

\bibitem[Pumain et~al., 2006]{pumain2006evolutionary}
Pumain, D., Paulus, F., Vacchiani-Marcuzzo, C., and Lobo, J. (2006).
\newblock An evolutionary theory for interpreting urban scaling laws.
\newblock {\em Cybergeo: European Journal of Geography}.

\bibitem[Raimbault, 2020]{raimbault2020model}
Raimbault, J. (2020).
\newblock A model of urban evolution based on innovation diffusion.
\newblock In {\em Artificial Life Conference Proceedings}, pages 500--508. MIT
  Press.

\bibitem[Raimbault et~al., 2019]{raimbault2019space}
Raimbault, J., Cottineau, C., Le~Texier, M., Le~Nechet, F., and Reuillon, R.
  (2019).
\newblock Space matters: Extending sensitivity analysis to initial spatial
  conditions in geosimulation models.
\newblock {\em Journal of Artificial Societies and Social Simulation}, 22(4).

\bibitem[Reuillon et~al., 2013]{reuillon2013openmole}
Reuillon, R., Leclaire, M., and Rey-Coyrehourcq, S. (2013).
\newblock Openmole, a workflow engine specifically tailored for the distributed
  exploration of simulation models.
\newblock {\em Future Generation Computer Systems}, 29(8):1981--1990.

\bibitem[Rozenblat and Pumain, 2018]{rozenblat2018conclusion}
Rozenblat, C. and Pumain, D. (2018).
\newblock Conclusion: Toward a methodology for multi-scalar urban system
  policies.
\newblock {\em International and Transnational Perspectives on Urban Systems},
  page 385.

\bibitem[Schmitt et~al., 2015]{schmitt2015half}
Schmitt, C., Rey-Coyrehourcq, S., Reuillon, R., and Pumain, D. (2015).
\newblock Half a billion simulations: Evolutionary algorithms and distributed
  computing for calibrating the simpoplocal geographical model.
\newblock {\em Environment and Planning B: Planning and Design},
  42(2):300--315.

\bibitem[Sharma and Mathew, 2011]{sharma2011multiobjective}
Sharma, S. and Mathew, T.~V. (2011).
\newblock Multiobjective network design for emission and travel-time trade-off
  for a sustainable large urban transportation network.
\newblock {\em Environment and Planning B: Planning and Design},
  38(3):520--538.

\bibitem[Singell, 1974]{singell1974optimum}
Singell, L.~D. (1974).
\newblock Optimum city size: Some thoughts on theory and policy.
\newblock {\em Land Economics}, 50(3):207--212.

\bibitem[Su et~al., 2017]{su2017density}
Su, H., Wei, H., and Zhao, J. (2017).
\newblock Density effect and optimum density of the urban population in china.
\newblock {\em Urban Studies}, 54(7):1760--1777.

\bibitem[Vigui{\'e} and Hallegatte, 2012]{viguie2012trade}
Vigui{\'e}, V. and Hallegatte, S. (2012).
\newblock Trade-offs and synergies in urban climate policies.
\newblock {\em Nature Climate Change}, 2(5):334--337.

\end{thebibliography}
\end{document}